\documentclass[preprint,5p,times,authoryear]{elsarticle}

\usepackage{amsmath}
\usepackage{caption}
\usepackage{array}
\usepackage{float}
\usepackage{mathtools}
\usepackage{tabularx}
\usepackage[usenames,dvipsnames]{xcolor}
\definecolor{forceblue}{rgb}{0.2235, 0.4157, 0.6941}
\usepackage[colorlinks]{hyperref}
\AtBeginDocument{%
	\hypersetup{
		colorlinks = true,
		citecolor=forceblue,
		linkcolor=forceblue,   
		urlcolor=forceblue,
		anchorcolor = forceblue}}
\usepackage{amsmath}
\usepackage{caption}
\usepackage{array}

\makeatletter
\def\ps@pprintTitle{%
	\let\@oddhead\@empty
	\let\@evenhead\@empty
	\def\@oddfoot{\footnotesize\itshape
		Preprint published in \ifx\@journal\@empty Elsevier
		\else\@journal\fi\hfill}
	\let\@evenfoot\@oddfoot
}
\makeatother

\journal{Journal of Wind Engineering and Industrial Aerodynamics 253 (2024) 105848 (\href{https://doi.org/10.1016/j.jweia.2024.105848}{doi.org/10.1016/j.jweia.2024.105848})}

\DeclareMathAlphabet{\mathcal}{OMS}{cmsy}{m}{n}
\SetMathAlphabet{\mathcal}{bold}{OMS}{cmsy}{b}{n}

\begin{document}
\begin{frontmatter}

\title{Data-driven Aeroelastic Analyses of Structures in Turbulent Wind Conditions using Enhanced Gaussian Processes with Aerodynamic Priors}
\author[a]{Igor Kavrakov\corref{cor}}
\cortext[cor]{Corresponding author.}
\ead{ik380@cam.ac.uk}
\author[b]{Guido Morgenthal}
\author[a]{Allan McRobie}
\address[a]{Department of Engineering, University of Cambridge, JJ Thomson Avenue 7a, Cambridge CB3 0FA, United Kingdom}
\address[b]{Chair of Modelling and Simulation of Structures, Bauhaus University Weimar, Marienstr. 13, Weimar 99423, Germany}

\begin{abstract}
Recent advancements in data-driven aeroelasticity have been driven by the wealth of data available in the wind engineering practice, especially in modeling aerodynamic forces. Despite progress, challenges persist in addressing free-stream turbulence and incorporating physics knowledge into data-driven aerodynamic force models. This paper presents a hybrid Gaussian Process (GPs) methodology for non-linear modeling of aerodynamic forces induced by gusts and motion on bluff bodies. Building on a recently developed GP model of the motion-induced forces, we formulate a hybrid GP aerodynamic force model that incorporates both gust- and motion-induced angles of attack as exogenous inputs, alongside a semi-analytical quasi-steady (QS) model as a physics-based prior knowledge. In this manner, the GP model incorporates the absent physics of the QS model, and the non-dimensional hybrid formulation enhances its appeal from an aerodynamic perspective. We devise a training procedure that leverages simultaneous input signals of gust angles, based on random free-stream turbulence, and motion angles, based on random broadband signals. We verify the methodology through analytical linear aerodynamics of a flat plate and non-linear aerodynamics of a bridge deck using Computational Fluid Dynamics (CFD). The standout feature of the presented methodology is its applicability for aeroelastic buffeting analyses, showcasing robustness when handling broadband excitation. Importantly, the non-linear hybrid model preserves its capability to capture higher-order harmonics in the motion-induced forces and remains applicable for flutter analysis, while incorporating both motion and gust angles as input. Applications of the methodology are anticipated in the aeroelastic analysis and monitoring of slender line-like structures.

\end{abstract}
\begin{keyword}
Gaussian Processes \sep Data-driven \sep Structural Aerodynamics \sep Aeroelasticity \sep Machine Learning \sep Buffeting \sep Flutter
\end{keyword}
\end{frontmatter}

\section{Introduction} \label{sec:Introduction}

Slender, line-like structures, such as long-span bridges and tall towers, are prone to wind-induced vibrations. Their sharp-edged cross sections characterize these structures as bluff bodies with complicated aerodynamics, particularly at high Reynolds numbers, where free-stream turbulence can lead to massive flow separation~\citep{bearmanEffectFreeStream1983}. The resulting gust- and motion-induced aerodynamic forces are often non-linear and non-Gaussian, making them a relevant topic for extensive research~\citep{kareemEmergingFrontiersWind2020,abbasMethodsFlutterStability2017}.\par
 
The aerodynamic forces acting on bluff bodies are typically represented using semi-analytical, white-box aerodynamic models that rest on the linear aerodynamic principles of a flat plate and aerodynamic coefficients obtained from wind tunnel testing or Computational Fluid Dynamics (CFD). Over the years, significant progress in modeling has been made, starting from linear unsteady models~\citep{scanlanMotionrelatedBodyforceFunctions2000,chenAdvancesModelingAerodynamic2002,dianaIABSETaskGroup2022}, evolving into hybrid models that utilize both linear unsteady and non-linear quasi-steady approaches~\citep{dianaNonlinearMethodCompute2020,barniBuffetingResponseSuspension2022}, and incorporating Volterra series as gray-box models~\citep{wuNonlinearAnalysisFramework2015,skyvulstadUseLaguerrianExpansion2021}. However, semi-analytical models are constrained by their mathematical constructions~\citep{kavrakovCategoricalPerspectiveAerodynamic2019}, thus, they are limited in terms of capturing non-linear aerodynamics and can be prohibitive due to the significant number of experiments required to extract their inherent parameters.\par

In the past decade, data-driven models have garnered significant attention in aerodynamics and aeroelasticity (see, e.g., \cite{kouDatadrivenModelingUnsteady2021} for a review). These models rely on learning a latent input-output function based on training data from CFD or experiments, which can then be used for prediction. Compared to semi-analytical aerodynamic models, data-driven models possess a broader range of mathematical properties. They can capture a wide array of dynamical properties, such as higher-order harmonics and fading fluid memory \citep{schoukensNonlinearSystemIdentification2019}.\par

The existing data-driven models in structural aerodynamics primarily focus on modeling motion-induced forces and predominantly rely on artificial neural networks \citep{abbasPredictionAeroelasticResponse2020,liNonlinearUnsteadyBridge2020,wuModelingHystereticNonlinear2011}. In a recent study, \cite{kavrakovDatadrivenAerodynamicAnalysis2022} employed Gaussian Processes (GPs) to develop a non-linear finite impulse response (GP-NFIR) model for motion-induced forces. Being non-parametric, GPs can approximate a wide class of non-linear functions with only a few hyperparameters inferred in a learning procedure that handles overfitting automatically~\citep{rasmussenGaussianProcessesMachine2006} and offer the possibility to introduce physics-based knowledge through a prior mean function. Wind engineering data is typically noisy, either due to measurement noise or inherent CFD noise; thus, GPs can be particularly suitable for modeling aerodynamic forces as they handle observation noise naturally. Generally, data-driven models have proven capable of simulating non-linear motion-induced aerodynamics, including coupled flutter and post-flutter limit cycle oscillations (LCOs). \par

\begin{figure}[!t]
	\centering
	\includegraphics[clip,height=4cm]{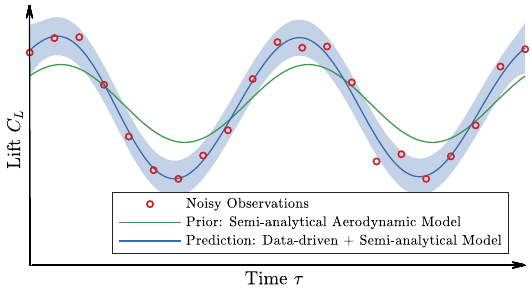} 
	\caption{Concept: The noisy observations of the lift are obtained through forced excitation tests (CFD or, potentially, experiments). The semi-analytical aerodynamic model (green) is a prior mean to the GP model. The GP model is trained on the noisy data (red circles), and can be used for prediction of both mean (blue line) and confidence interval (shaded region) since it is a probabilistic regression.}
	\label{fig:Concept}
\end{figure} 

Despite progress, including gust- and motion-induced contributions in data-driven aerodynamic force models remains an unresolved issue. This question is crucial, given that the buffeting response typically governs structural design. Moreover, relying solely on data-driven inference may compromise robustness; thus, integrating prior physics-based knowledge is advocated to enhance the reliability and performance of data-driven frameworks~\citep{levineFrameworkMachineLearning2022}.\par

In this study, we introduce a hybrid Gaussian Processes (GP) methodology for buffeting and flutter analyses of bluff bodies. Expanding on the GP-NFIR aerodynamic force model by \cite{kavrakovDatadrivenAerodynamicAnalysis2022}, we extend the model to include gust- alongside motion-induced angles of attack. Furthermore, we enhance the GP model by integrating the semi-analytical quasi-steady (QS) model as a physics-based prior mean (cf. Fig.~\ref{fig:Concept}). This results in a non-dimensional (in an aerodynamic sense) and non-parametric (in a statistical sense) hybrid representation of aerodynamic forces. To generate training data, we devise an intuitive approach for design of training experiments. Subsequently, we verify the methodology using flat plate linear aerodynamics, and finally, apply it to conduct flutter and buffeting analyses of a 2D section of the Great Belt Bridge's deck, showcasing its non-linear capabilities.

\section{Enhanced GP Model with Aerodynamic Priors}

\subsection{Formulation}

Consider the equation of motion of a two-degree-of-freedom system (cf. Fig.~\ref{fig:WS}) given as:
\begin{equation}\label{eq:EqMot}
		\begin{aligned}
			m_h\ddot{h}+c_h\dot{h}+k_hh=L,\\
			m_\alpha\ddot{\alpha}+c_\alpha\dot{\alpha}+k_\alpha \alpha=M,
		\end{aligned}
\end{equation}
where $h=h(t)$, $\alpha=\alpha(t)$ are the displacements; $m_h$, $m_\alpha$ are the masses; $c_h$, $c_\alpha$ are the damping coefficients; $k_h$, $k_\alpha$ are the stiffness coefficients in vertical and rotational direction, respectively. The dot $\dot{()}$ denotes derivative with respect to time $t$. The fluctuating lift and the moment forces are 

\begin{equation}\label{eq:forces}
L(t)=-\frac{1}{2}\rho U^2BC_L(t),\hspace*{0.5cm}	M=\frac{1}{2}\rho U^2B^2C_M(t),\\
\end{equation}

where $\rho=$1.2 kg/m$^3$ is the air density; $U$ is the mean wind speed; and, $B$ is the chord width. The aerodynamic forces depend on the motion and incoming free-stream turbulence, the latter described through the $u=u(t)$ lateral and $w=w(t)$ vertical turbulent components. \par
\begin{figure}[!t]
	\centering
	\includegraphics[clip]{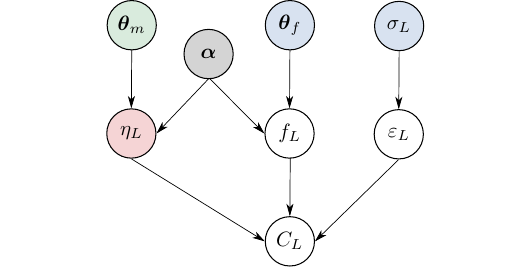} 
	\caption{Graphical model of the GP-NFIR model of the lift $C_L$ for non-dimensional angles of attack $\boldsymbol{\alpha}$: $\eta_L$ is the aerodynamic model as a physics-based prior mean with parameters $\boldsymbol{\theta}_m$; $f_L$ is the probabilistic GP model with hyperparameters $\boldsymbol{\theta}_f$; $\varepsilon_L$ is the Gaussian noise with variance $\boldsymbol{\sigma}_L$. The gray circle represents deterministic variables; the white circles represent random variables (or a collection of); the blue circles represent unknown parameters; the green circle represent known parameters and the red circle represents a deterministic variable.}
	\label{fig:ProbModel}
\end{figure}

We formulate the fluctuating lift $C_L(t)$ and moment $C_M(t)$ coefficients as a separate probabilistic input/output models, enhanced with aerodynamic priors (cf. Fig.~\ref{fig:ProbModel}). Considering the lift as an example (moment follows similar approach), the model is described as a sum of three components: 

\begin{figure}[!t]
	\centering
	\includegraphics[clip]{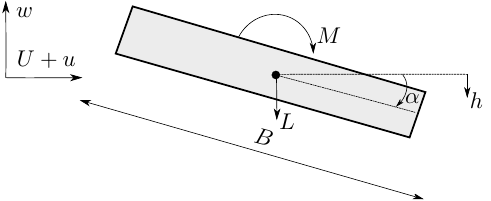} 
	\caption{Coordinate system of aerodynamic forces acting on a bluff body.}
	\label{fig:WS}
\end{figure}
\begin{equation}~\label{eq:CL}
	\begin{aligned}
	C_L(t)=\eta_L(t)+f_L(t)+\varepsilon_L(t),
\end{aligned}
\end{equation}

where $\eta_L$ is the semi-analytical aerodynamic model as a deterministic physics-based prior with model parameters $\boldsymbol{\theta}_m$; $f_L$ is a probabilistic data-driven model with hyperparameters $\boldsymbol{\theta}_f$; and, $\varepsilon_L\sim\mathcal{N}(0,\sigma_L^2)$ is the observation noise, assumed to be Gaussian with variance $\sigma_L^2$. The model parameters $\boldsymbol{\theta}_m$ are inherently linked to the choice of the semi-analytical model (e.g., static aerodynamic coefficients) and are obtained through specific CFD or wind tunnel experiments (e.g., static aerodynamic tests). The hyperparameters $\boldsymbol{\theta}_f$ related to the data-driven model are learned based on a training dataset obtained from experiments that differ from those used for $\boldsymbol{\theta}_m$. We propose a bespoke procedure for training experiments in Sec.~\ref{sec:DesignTrainingExp}.\par
The first component $\eta_L$ describes our knowledge of the aerodynamic system, i.e. a semi-analytical aerodynamic model. Although any model can be used as an aerodynamic prior mean in the methodology, we employ the QS model $\eta_L = \eta_L(\alpha_e, \boldsymbol{\theta}_m; t)$ because it can account for QS aerodynamic non-linearity~\citep{kavrakovSynergisticStudyCFD2018}. Moreover, its mathematical construction is simple, as it depends on the instantaneous effective angle of attack $\alpha_e$, making it computationally efficient to integrate within our framework. The QS model is formulated as:

\begin{equation}~\label{eq:QS}
		\begin{aligned}
			\eta_L:=C_{LQS}(t)=\tilde{C}_{L}(\alpha_e(t)),\hspace*{0.2cm}		\alpha_e(t)=\alpha+\arctan\left(\frac{w+\dot{h}+n_LB\dot{\alpha}}{U+u}\right).
		\end{aligned}
\end{equation}
where the fluctuating QS lift $C_{LQS}(t)$ is obtained by interpolating the mean static lift coefficient $\tilde{C}_L(\alpha)$ (obtained through static aerodynamic tests) at the instantaneous effective angle of attack $\alpha_e$. 
In the case of the QS model, the aerodynamic model parameters are denoted as $\boldsymbol{\theta}_m=(\tilde{C}_L, n_L)$, where $n_L$ represents the aerodynamic center, typically obtained from flutter derivatives. We determine the parameters $\boldsymbol{\theta}_m$ through separate CFD or wind tunnel experiments, resulting in a deterministic output $\eta_L$ (cf. Fig.~\ref{fig:ProbModel}). We note that the standard QS model~\citep{chenAdvancesModelingAerodynamic2002} is based on the resultant wind speed instead of the mean wind speed $U$ in~\eqref{eq:forces}. Nevertheless, we use the mean wind speed $U$ for simplicity in our formulation, as any resulting discrepancy is captured by the data-driven part of the model.\par

The second component, the data-driven model $f_L$, captures all effects that are intractable for the QS model, such as non-linear fluid memory. Two characteristics are important when constructing the data-driven model: first, it needs to be non-dimensional for easy translation between scales; and second, it needs to be a class of functions that is sufficiently flexible to capture non-linear phenomena. Extending the formulation by~\cite{kavrakovDatadrivenAerodynamicAnalysis2022} to include the effect of turbulence, the model is formulated with the motion and gust angles of attack as exogenous inputs:
\begin{equation}\label{eq:Form}
f_L=f_L(\alpha_h,\alpha_h^\prime,\alpha_a,\alpha_a^\prime,\alpha_u,\alpha_u^\prime,\alpha_w,\alpha_w^\prime,\boldsymbol{\theta}_f;\tau,\tau-\tau_m.)\end{equation}
The motion, gust, and effective angles of attack can be represented in a non-dimensional form:
\begin{equation}\label{eq:non-dim}
		\begin{aligned}
		&\begin{aligned}
			\alpha_a&=\alpha&&\alpha_a^\prime=\frac{B}{U}\dot{\alpha_a}, &&&h^\prime=\frac{B}{U}{\dot{h}}\\
			\alpha_h&=\arctan\left(\frac{h^\prime}{B}\right),&&\alpha_h^\prime=\frac{1}{\displaystyle 1+\left(\frac{h^\prime}{B}\right)^2}\frac{h^{\prime\prime}}{B}\\
			\alpha_u&=\arctan\left(\frac{u}{U}\right),&&\alpha_u^\prime=\frac{1}{\displaystyle1+\left(\frac{u}{U}\right)^2}\frac{u^\prime}{U}\\
			\alpha_w&=\arctan\left(\frac{w}{U}\right),&&\alpha_w^\prime=\frac{1}{\displaystyle1+\left(\frac{w}{U}\right)^2}\frac{w^\prime}{U}
		\end{aligned}\\
			&\alpha_e=\alpha_a+\arctan\left(\frac{\tan\alpha_w+\tan\alpha_h+n_L\alpha_a^\prime}{1+\tan\alpha_u}\right)\end{aligned}
\end{equation}
where the $()^\prime$ denotes derivative with respect to non-dimensional time $\tau=tU/B$, considering $\text{d}t/\text{d}\tau=B/U$. The system memory, necessary in aerodynamics, is considered in~\eqref{eq:Form} through lagged input up to time $\tau-\tau_m$. \par 
The latent mapping function $f_L$ is modeled through a GP as a Bayesian probabilistic regression~\citep{rasmussenGaussianProcessesMachine2006}. By placing an infinite-dimensional space of functions as a prior on $f_L$, we obtain:
\begin{equation}\label{eq:GPprior}
f_L\sim \mathcal{GP}(0,k_L),
\end{equation}
where $k_L$ is the covariance function that depends on certain hyperparameters $\boldsymbol{\theta}_f$, which, for the time being, we consider fixed. The data-driven model $f_L$ relies on a discrete training dataset obtained from measurements (CFD or wind tunnel). The discrete noisy observations of the fluctuating lift coefficient in~\eqref{eq:CL} are available in a vector $\boldsymbol{C}_L\in\mathbb{R}^{N_s\times1}$, where $N_s$ is the number of time steps. The training dataset is then represented as $\left\{\boldsymbol{\alpha},\boldsymbol{C}_L\right\}$, wherein the input matrix $\boldsymbol{\alpha}\in\mathbb{N}^{N_s\times8+4S}$ is comprised of stacked time histories of the effective angles of attack in~\eqref{eq:non-dim}, which at each time step $i$ has the form:
\begin{equation}\label{eq:Input}
		\begin{aligned}
			\boldsymbol{\alpha}_i=(&\alpha_{h,i}^\prime,\alpha_{a,i}^\prime,\alpha_{u,i}^\prime,\alpha_{w,i}^\prime,\alpha_{h,i},\alpha_{a,i},\alpha_{u,i},\alpha_{w,i},\dots\\
			&\alpha_{h,i-S},\alpha_{a,i-S},\alpha_{u,i-S},\alpha_{w,i-S}),
		\end{aligned}
\end{equation}
where $S=\tau_m/\Delta\tau$ is the number of lagged (past) input for $\Delta \tau$ being the non-dimensional time step. The data-driven prior~\eqref{eq:GPprior} then becomes a multivariate Gaussian distribution: 
\begin{equation}~\label{eq:PriorDisc}
		\begin{aligned}
			p(\boldsymbol{f}_L|\boldsymbol{\alpha})=\mathcal{N} (\boldsymbol{0},\boldsymbol{K}_L),
		\end{aligned}
\end{equation}
where $\boldsymbol{K}_L=k_L(\boldsymbol{\alpha},\boldsymbol{\alpha})$ is the covariance matrix of size $\boldsymbol{K}\in\mathbb{N}^{N_s\times N_s}$. The covariance matrix is obtained based on the covariance function $k_L$, which is dependent on the task at hand. In this study, we employ the exponential kernel with Automatic Relevance Detection (ARD) (cf.~\ref{App:CovFunc}). This kernel is non-linear and infinitely differentiable, making it an appropriate choice for non-linear dynamical systems since it is smooth in time \citep{frigolaVariationalGaussianProcess2014}. Moreover, each dimension is scaled independently due to the anisotropic ARD property, making it particularly useful for purposes in structural aerodynamics due to the effect of fluid memory associated with each lag term in the input vector (see~\eqref{eq:Input}). It is noted that the kernel choice warrants further investigation or use of automatic construction of kernel functions \citep{calandraManifoldGaussianProcesses2016,duvenaudAutomaticModelConstruction2014}. \par 
The likelihood of the discrete lift observations $\boldsymbol{C}_L$ for (deterministic) training input $\boldsymbol{\alpha}$ in~\eqref{eq:CL} can be described using the QS prior mean in~\eqref{eq:QS}, through its discrete predictions $\boldsymbol{C}_{LQS}=\tilde{C}_L(\boldsymbol{\alpha})\in\mathbb{R}^{N_s\times1}$, the data-driven prior in~\eqref{eq:PriorDisc}, and the Gaussian noise $\boldsymbol{\varepsilon}\sim\mathcal{N}(\boldsymbol{0},\sigma_L^2\boldsymbol{I})$. As both the noise and the data-driven prior are Gaussian, the likelihood is also Gaussian:
\begin{equation}
		p(\boldsymbol{C}_L|\boldsymbol{f},\boldsymbol{\alpha})=\mathcal{N}(\boldsymbol{C}_{LQS}+\boldsymbol{f}_L,\sigma_L^2\boldsymbol{I}),
\end{equation}
from which the data-driven prior distribution $p(\boldsymbol{f}_L)$ can be marginalized ~\citep{rasmussenGaussianProcessesMachine2006}, yielding the marginal likelihood
\begin{equation}\label{eq:ML}
	p(\boldsymbol{C}_L|\boldsymbol{\alpha})=\mathcal{N}(\boldsymbol{C}_{LQS},\boldsymbol{K}_L+\sigma_L^2\boldsymbol{I}).
\end{equation}\par
The marginal likelihood describes how well the presented model in~\eqref{eq:CL} describes the training data: $\boldsymbol{C}_{LQS}$ is the semi-analytical part, $\boldsymbol{K}_L$ is the data-driven part and $\sigma_L^2\boldsymbol{I}$ is the observation noise. \par

\subsection{Training}
The kernel hyperparameters $\boldsymbol{\theta}_f$ and noise variance $\boldsymbol{\sigma}_L^2$ should be inferred from the training data. To determine them, we maximize the logarithm of the marginal likelihood (cf.~\eqref{eq:ML}) with respect to an extended hyperparameter vector $\boldsymbol{\theta} = (\boldsymbol{\theta}_f, \boldsymbol{\sigma}_L^2)$, which provides point estimates. Alternatively, the hyperparameters can be treated in a fully probabilistic manner in a Bayesian setting, i.e., they can be assigned a prior distribution and their posterior obtained using Monte Carlo methods.\par
Considering the hyperparameters as optimisation variables, the optimisation task is defined as:
\begin{equation}\label{eq:Minimisation}
		\arg\max_{\boldsymbol{\theta}} \log p(\boldsymbol{C}_L|\boldsymbol{\alpha};\boldsymbol{\theta}).
\end{equation}\par
The objective function is tractable since the marginal likelihood~\eqref{eq:ML} is Gaussian. Taking the logarithm of the marginal likelihood yields~\citep{rasmussenGaussianProcessesMachine2006}
\begin{equation}\label{eq:logLike} 
		\begin{aligned}
			\log p(\boldsymbol{C}_L|\boldsymbol{\alpha};\boldsymbol{\theta})&=-\frac{1}{2}(\boldsymbol{C}_L-\boldsymbol{C}_{LQS})^T(\boldsymbol{K}_{L}+\sigma_L^2\boldsymbol{I})^{-1}(\boldsymbol{C}_L-\boldsymbol{C}_{LQS})\\ &-\frac{1}{2}\log\left|\boldsymbol{K}_{L}+\sigma_L^2\boldsymbol{I}\right|-\frac{N_p}{2}\log 2\pi.
		\end{aligned}
\end{equation}\par
The marginal likelihood is a convex function, which means that a gradient-based optimizer with analytical derivatives can be used to obtain point estimates of $\boldsymbol{\theta}$. It also adheres to Occam's razor by providing a trade-off between the data fit and statistical model complexity; thereby, preventing model overfit~\citep{rasmussenOccamRazor2000}.\par
The model parameter $\boldsymbol{\theta}_m$ can be also treated similarly to the kernel hyperparameters $\boldsymbol{\theta}_f$, with initial values based on separate wind tunnel experiments (e.g., static tests). Thus, they can be inferred as a point estimate during the maximization of the log-likelihood or be treated fully probabilistically and inferred through Monte Carlo optimization with a prescribed prior distribution.

\subsection{Prediction}
Having determined the hyperparameters $\boldsymbol{\theta}$, we are interested to leverage the training data $\left\{\boldsymbol{\alpha},\boldsymbol{C}_L\right\}$ through~\eqref{eq:ML} to obtain the predictive posterior distribution $p(\boldsymbol{C}_L^*)$ at (deterministic) prediction inputs $\boldsymbol{\alpha}^*$. This can be obtained through the joint prior distribution
\begin{equation}~\label{eq:Joint}
	p(\boldsymbol{C}_L,	\boldsymbol{C}_L^*|\boldsymbol{\alpha}^*,\boldsymbol{\alpha})= 
	\mathcal{N}\left(\begin{bmatrix}
		\boldsymbol{C}_{LQS}\\
		\boldsymbol{C}_{LQS}^*
	\end{bmatrix},\begin{bmatrix}
		\boldsymbol{K}_L+\sigma_L^2\boldsymbol{I} & \boldsymbol{K}_{L^*}\\
		\boldsymbol{K}_{L^*}^T & \boldsymbol{K}_{LL^*}
	\end{bmatrix}\right),
\end{equation}
where $\boldsymbol{K}_{L^*}= k_L(\boldsymbol{\alpha}^*,\boldsymbol{\alpha}^*)$, $\boldsymbol{K}_{LL^*}= k_L(\boldsymbol{\alpha},\boldsymbol{\alpha}^*)$ are covariance matrices obtained based on the kernel~\eqref{eq:Kernel}, and $\boldsymbol{C}_{LQS}^*=\tilde{C}_L(\boldsymbol{\alpha}^*)$ is the predictive QS mean. Conditioning the joint distribution on the lift observations $\boldsymbol{C}_L$ yields the posterior distribution:
\begin{equation}\label{eq:PredDist}
	p(\boldsymbol{C}_L^*|\boldsymbol{C}_L,\boldsymbol{\alpha}^*,\boldsymbol{\alpha})=\mathcal{N}(\boldsymbol{m}_L^*,\boldsymbol{K}_L^*),
\end{equation}
with posterior mean $\boldsymbol{m}_L^*$ and covariance $\boldsymbol{K}_L^*$:
\begin{equation}
		\begin{aligned}\label{eq:PredMean}
			\boldsymbol{m}_{L}^*&=\boldsymbol{C}_{LQS}^*+\boldsymbol{K}_{LL^*}^T(\boldsymbol{K}_{L}+\sigma_L^2\boldsymbol{I})^{-1}(\boldsymbol{C}_L-\boldsymbol{C}_{LQS}),\\
			\boldsymbol{K}_{L}^*&=\boldsymbol{K}_{L^*}-\boldsymbol{K}_{LL^*}^T(\boldsymbol{K}_{L}+\sigma_L^2\boldsymbol{I})^{-1}\boldsymbol{K}_{LL^*}.
		\end{aligned}
\end{equation}\par
Thus, the predictive posterior $p(\boldsymbol{C}_L^*)$ is a multivariate Gaussian distribution, completely described by its posterior mean $\boldsymbol{m}_{L}^*$ and covariance $\boldsymbol{K}_{L}^*$. These are obtained based on the training input $\boldsymbol{\alpha}$, training output $\boldsymbol{C}_L$, training QS mean $\boldsymbol{C}_{LQS}$, predictive QS mean $\boldsymbol{C}_{LQS}^*$ and prediction input $\boldsymbol{\alpha}^*$.\par
This concludes the formulation of the GP-NFIR model with a QS prior mean. The model is completely non-dimensional depending on the angles of attack (cf.~\eqref{eq:non-dim}), making it suitable for applications in structural aerodynamics. Including analytical priors and capturing the model error by a data-driven model is shown to require less data and provides more efficient inference~\citep{levineFrameworkMachineLearning2022}.\par 
The presented hybrid GP model is sufficiently expressive to capture non-stationary and non-linear forces with respect to time. The effective angles~\eqref{eq:non-dim} and their memory effectively warp the time variable~\citep{mackay1998introduction}; thus, there are no restraints on the input motion or gust. However, it is noted that all data-driven models are only as good as their training dataset; therefore, an appropriate design of experiments is a prerequisite for accurate prediction, as discussed in Sec.~\ref{sec:DesignTrainingExp}.

\subsection{Framework}

In practice, we train our model for a training dataset $\left\{\boldsymbol{\alpha},\boldsymbol{C}_L\right\}$ (see Sec.~\ref{sec:DesignTrainingExp}) by optimizing the hyperparameters $\boldsymbol{\theta}$ through~\eqref{eq:Minimisation} and~\eqref{eq:logLike}. The aerodynamic prior mean $\boldsymbol{C}_{LQS}$ is the QS model in~\eqref{eq:QS}, obtained by interpolating the static wind coefficients $\tilde{C}_L$ at the training input $\boldsymbol{\alpha}$ (non-dimensional effective angle of attack - cf.~\eqref{eq:non-dim}). The covariance matrix $\boldsymbol{K}_L$ is constructed using the covariance function in~\eqref{eq:Kernel} and the input matrix $\boldsymbol{\alpha}$, where each dimension is scaled with the corresponding length scales~\eqref{eq:Kernel2}. These length scales are part of the hyperparameters $\boldsymbol{\theta}$ learned through minimizing the marginal likelihood in~\eqref{eq:Minimisation}. To improve efficiency, we invert the nosiy covariance matrix $\boldsymbol{K}_L+\sigma_L^2\boldsymbol{I}$ in~\eqref{eq:logLike} using Cholesky decomposition (see Algorithm 2.1 of ~\cite{rasmussenGaussianProcessesMachine2006}). The training data can be sub-sampled with a factor of $F$ as a strategy to reduce computational cost of $\mathcal{O}(n^3)$ during training \citep{quinonero-candelaUnifyingViewSparse2005}. \par 
During prediction, we obtain the posterior distribution $p(\boldsymbol{C}_L^*)$~\eqref{eq:PredDist} using the posterior mean and covariances ~\eqref{eq:PredMean} at prediction input $\boldsymbol{\alpha}^*$. It can be observed that the posterior mean consists of the aerodynamic QS prior mean $\boldsymbol{C}_{LQS}^*$ at prediction points $\boldsymbol{\alpha}^*$, and an additional term related to the covariance function that captures the model inadequacy of the QS model with respect to the training data. The full training data is used when conditioning in the prediction. The prediction can be either multi-step ahead, i.e. forced excitation where the predictive angle $\boldsymbol{\alpha}^*$ is prescribed at all times, or dynamic one-step ahead, i.e. dynamic simulation where the predictive angle $\boldsymbol{\alpha}^*$ is provided one step ahead from time integration of the coupled equation of motion~\eqref{eq:EqMot} (e.g. Newmark-Beta scheme~\citep{cloughDynamicsStructures1975}). In the case of a dynamic simulation, the predictive posterior mean ${m}_{L,i}^*$ at step $t_i$ is obtained based on the input matrix $\boldsymbol{\alpha}_i^*$ and fed forward in the time integration to obtain the displacements at the time step $t_{i+1}$. These displacements are then used to form the input matrix $\boldsymbol{\alpha}_{i+1}^*$ and predict the mean force ${m}_{L,i+1}^*$, and so forth. It is noted that a fully probabilistic dynamic analysis would need to include Monte Carlo sampling at each step of the predictive posterior $p(\boldsymbol{C}_L^*)$. Similar expressions can be obtained for the moment, with the training data $\left\{\boldsymbol{\alpha},\boldsymbol{C}_M\right\}$, static moment coefficient $\tilde{C}_M$, and moment aerodynamic center $n_M$.

\section{Design of Training Experiments}\label{sec:DesignTrainingExp}

In this section, we outline the process of generating training data for the hybrid GP model. The training data should represent the archetype of the model's prediction space \citep{schoukensNonlinearSystemIdentification2019}. Our model is designed for buffeting and flutter analyses of bluff bodies. Consequently, the gust input is simulated using stationary free-stream turbulence, while the motion input is generated through broadband random excitation. Important parameters for the training signal include gust and motion amplitudes, as well as their frequency content.\par
Due to the non-linear nature of the model, it is important that gust and motion act concurrently in at least one sample of the training time history. In simpler terms, the training data is created to simulate bluff bodies undergoing forced vibrations while immersed in free-stream turbulence.

\begin{figure*}[!t]
	\centering
	\includegraphics[clip]{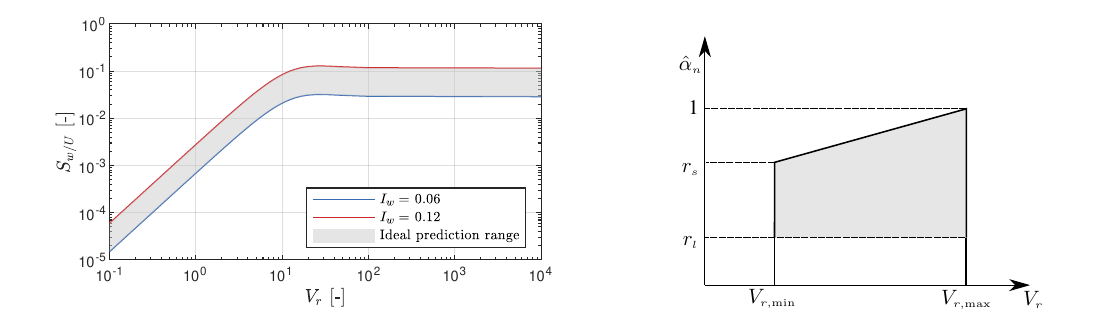} 
	\caption{Generation of training data set. von K\'arm\'an spectrum of the non-dimensional vertical gust amplitude $\alpha_w\approx w/U$ with $\tau_w=2$ and $r_{bw}=1$ (left). Fourier amplitudes of a signal designed for training with the shaded area representing the sampling interval for the Fourier amplitudes (right).}
	\label{fig:Karman}
\end{figure*}

\begin{table*}[t]
	\centering
	\footnotesize
		\begin{tabularx}{0.93\textwidth}{l | c c c c c |c c c c c c |c c c c c c}
		\hline
		
		\multicolumn{1}{c}{ }&\multicolumn{5}{c}{Basic Parameters} & \multicolumn{6}{c}{Motion Angles Parameters} &\multicolumn{6}{c}{Gust Angles Parameters}  \\[2pt]
		Sec. & B &$\Delta\tau$ & $\tau$  & $S$& $F$&  $V_{r,\min}$  & $ V_{r,\max}$ & $\sigma_{\alpha_h}$ & $\sigma_{\alpha_a}$ & $r_l$ & $r_s$ & $I_u$  & $I_w$  & $\tau_u$  &  $\tau_w$  & $r_{Bu}$ & $r_{Bw}$ \\[2pt]	
		&[m] & [-]  & [-] & [-]   & [-] & [-] & [-] & [deg] & [deg] & [-] & [-]& [\%] & [\%]& [-] & [-] & [-] & [-] \\[2pt]		
		\hline
		FP& 31 & 0.05 & 2$\times$280 & 200$^*$   & 3 & 2 & 14 & 0.5/2.5 & 0.5/2.5 & 0.05 & 1.0  & / & 3/6 & / & 2.00 & / & 2.00 \\[4pt]
		GB  & 31 & 0.05  & 2$\times$1000  & 200$^*$ & 5  & 2 &  16 & 2.5/15 & 2.5/15 & 0.05 & 0.2  & 6/12 & 6/12 & 2.70 & 1.35 & 1.75 & 	0.87 \\
		\hline
	\end{tabularx}
	\caption{Input and training parameters for the flat plate (FP) and Great Belt's deck section (GB): $\mathrm{B}$=chord width; $\Delta\tau$=reduced GP time step; $\tau$= training time of each signal; $S$=number of lags; $F$=sub-sampling factor; $V_{r,\min},V_{r,\max}$= minimum and maximum reduced velocities of motion training angles;  $\sigma_{\alpha_h}$,$\sigma_{\alpha_a}$=standard deviation of motion training angles; $r_l,r_s$= minimum and maximum relative Fourier amplitudes; $I_u,I_w$= turbulence intensities of wind fluctuations for training; $\tau_u,\tau_w$=characteristic times of wind training spectra; $r_{Bu},r_{Bw}$=turbulent length scales \textit{vs} chord width ratio. In case there are two training signals with low $\alpha^l$ and high $\alpha^h$ amplitudes, these are noted in the same table cell as $\sigma_{\alpha^l}/ \sigma_{\alpha^h}$. ($^*$ in case of the longitudinal velocity $u$, we use $S$=1 lag as the memory in this term is deemed negligent.)}
	\label{tab:NumParam}
\end{table*}

\subsection{Gust angles} 

In the case of the gust angles, one way to create an input training signal is by selecting the excitation similar to the structure of the atmospheric turbulence. To this end, it is beneficial to characterize the turbulent spectra based on the gust angles $\alpha_u$ and $\alpha_w$ and relate them to the structural dimension $B$. We begin with the von K\'arm\'an spectrum~\citep{vonkarmanProgressStatisticalTheory1948} for vertical turbulence as an illustrative example. It's worth noting that similar approach can be applied for other type of spectra, as they usually depend on the reduced frequency $fL_w/U$. The spectrum $S_w$ of the vertical fluctuations is given by:
\begin{equation}\label{eq:SpectraW}
	\frac{fS_{w}(f)}{\sigma_w^2}=\frac{4fL_w/U\left[1+755.2(fL_w/U)^2\right]}{\left[1+283.2(fL_w/U)^2\right]^{11/6}},
\end{equation}
where $f$ is the gust frequency, $L_w$ is the turbulent length scale and $\sigma_w^2$ is the variance of the vertical turbulence.\par
Consider the following parameters: 
\begin{equation}\label{eq:InputTrainingB}
\begin{aligned}
I_w=\frac{\sigma_w}{U}, \hspace*{0.5cm}	\tau_w=\frac{L_w}{U},\hspace*{0.5cm} r_{Bw}=\frac{L_w}{B}, \hspace*{0.5cm} V_r=\frac{U}{fB},\\
\end{aligned}
\end{equation}
where $I_w$ is the turbulence intensity; $\tau_w$ is a characteristic time required for a fluid particle to travel distance equivalent to the length scale; $r_{Bw}$ is the ratio between the length scale and characteristic dimension; and, $V_r$ is the reduced velocity, with respect to the gust frequency. Thus,~\eqref{eq:SpectraW} can be rewritten as the spectrum ${S_{w/U}}$ of the gust amplitude $w/U$ as:
\begin{equation}\label{eq:SpectraAW}
S_{w/U}(V_r)=\frac{4I_w^2\tau_w\left[1+755.2(r_{Bw}/V_r)^2\right]}{\left[1+283.2(r_{Bw}/V_r)^2\right]^{11/6}}.
\end{equation}
This non-dimensional expression is insightful as it depends on the reduced velocity $V_r$. Various properties of the aerodynamic forces can be characterized depending on $V_r$. For e.g., having $V_r\geq50$ usually means that the buffeting forces abide the QS assumption.\par 
In the case of small angles, the gust amplitude can be approximated as $\alpha_w\approx w/U$; thus, the turbulence intensity roughly corresponds to the standard deviation of $\alpha_w$. When selecting the gust amplitude, it is essential to consider its relevance to the intended application. Alternatively, a range of amplitudes could be explored by combining two training datasets with different amplitudes, denoted as $\alpha_w=(\alpha_{w1},\alpha_{w2})$. For instance, one dataset might include small gust angles, while the other involves large gust angles. The resulting model can then be applied to turbulence intensities roughly within these defined limits, as illustrated in Fig.~\ref{fig:Karman} (left).\par
If the training data is derived from CFD simulations, a straightforward approach involves scaling the standard deviation, $I_w$, while shifting the spectrum on the ordinate and keeping other non-dimensional parameters constant. For wind tunnel data, two key considerations arise: Firstly, obtaining a set of gust amplitudes can be challenging due to the inherent nature of turbulence~\citep{laroseDynamicActionGusty2002}. However, as long as the prediction range aligns with the region between the spectra of the training sets, satisfactory results can be expected. Secondly, it is important to note that the GP model presented here is based on the strip assumption and does not depend on the wave number in the longitudinal direction (see~\cite{kavrakovDeterminationComplexAerodynamic2019} for discussion). In cases of boundary-layer or grid-generated turbulence, this assumption may not hold unless the ratio between length-scale-to-structural-length is considered, though it is commonly valid on-site for line-like structures~\citep{massaroEffectThreedimensionalityAerodynamic2015}.\par 
Experiments to generate a training dataset could be designed using an Active Turbulence Generator~\citep{dianaNonlinearMethodCompute2020}, ensuring the retention of the strip assumption, although a larger amount of training data may be required in this scenario. Alternatively, a factor accounting for 3D effects in forces could be developed as proposed in~\cite{liThreedimensionalAerodynamicLift2023}. The specifics of the generation of gust angle training data in a wind tunnel represent a subject for further study.

\subsection{Motion angles}

Generating input training motion is significantly simpler since the frequency and amplitude content of the random excitation can be user-defined to an extent. We follow the procedure by \cite{kavrakovDatadrivenAerodynamicAnalysis2022} that we briefly revisit here.\par
Consider the vertical motion as an example. The frequency and amplitude content of the normalized angle $\alpha_{h,n}$ (with unit variance) are described through a random Fourier amplitude $\hat\alpha_{h,n}$ that follows a uniform distribution (see Fig.~\ref{fig:Karman}, right):
\begin{equation}
		\hat\alpha_{h,n}(V_r)\begin{cases}\sim\mathcal{U}[r_l,r_h],&\text{if}\ V_{r,\min}\leq V_r \leq V_{r,\max},\\=0,&\text{otherwise},\end{cases}
\end{equation}
where $V_r=U/(fB)$ is the reduced velocity, depending on the oscillation frequency $f$; $V_{r,\min}$ and $V_{r,\max}$ define the selected training range of reduced velocities; $r_l$ is the minimum Fourier amplitude (typically $r_l\approx0.05$); and,  $r_h=r_h(V_r)$ is the maximum Fourier amplitude. The latter is linearly dependent on the reduced velocity, such that, $r_h(V_{r,\min})=r_s$ and $r_h(V_{r,\max})=1$, for $r_l\leq r_s\leq 1$ (typically $r_s\approx0.2$).\par
Defining the frequency range in such a manner ensures, through $r_l$, that all frequencies in the range $\left[V_{r,\min}, V_{r,\max}\right]$ are excited. Moreover, the factor $r_h$ can constrain the relative amplitudes at high frequencies (i.e., low $V_r$) to mitigate practical issues, such as numerical instabilities in unresolved CFD simulations or mechanically infeasible experiments for forced vibrations.\par
The time-history of the angle $\alpha_{h,n}$ is obtained through the random Fourier transform: 
\begin{equation}\label{eq:InputTraining}
		\alpha_{h,n}(\tau)=\int_{-\infty}^{\infty}\hat{\alpha}_{h,n}(V_r)\exp\left(i\frac{2\pi}{V_r}+i\varphi\right)\text{d}V_r,
\end{equation}
where $i$ is the imaginary unit and $\varphi\sim\mathcal{U}[0,2\pi)$ is a random, uniformly distributed phase for each harmonic.\par 
Finally, the training signal $\alpha_h$ is scaled with the training (target) standard deviation $\sigma_{\alpha_h}$ as:
\begin{equation}\label{eq:InputTrainingScaling}
		\alpha_h=\alpha_{h,n}\frac{\sigma_{\alpha_{h}}}{\sigma_{\alpha_{h,n}}}.
\end{equation}\par 
As for the gust angles, two or more training signals with different amplitudes can be stacked together  $\alpha_w=(\alpha_{w1},\alpha_{w2})$ to cover both linear and non-linear oscillation range.


\begin{figure*}[!t]
	\centering
	\includegraphics[clip]{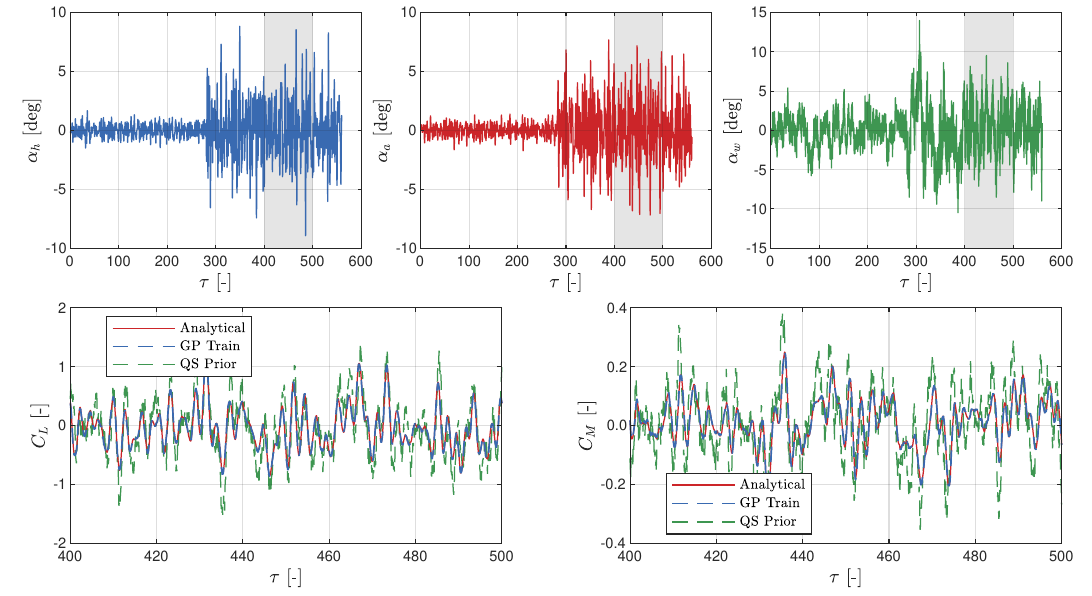} 
	\caption{Flat Plate - Forced Excitation (training): Input training signals of the vertical motion (top-left), rotation (top-centre) and gust (top-right) angles; sample training time histories of the aerodynamic coefficients (bottom) for the shaded region (top). The confidence interval for the wind coefficients is negligible.}
	\label{fig:FP_Train}
\end{figure*}

\section{Flat Plate}\label{sec:FundApp}

We verify the presented methodology based on the linear unsteady model for analytical flat plate aerodynamics. In its general form in the frequency domain, the linear unsteady aerodynamic model is expressed as follows~\citep{scanlanMotionrelatedBodyforceFunctions2000, davenportResponseSlenderLinelike1962}:
\begin{equation}\label{eq:FP_Forces}
\begin{aligned}
		L=\frac{1}{2}\rho U^2B&\Bigg( KH^*_1\frac{\dot{h}}{U}+KH^*_2\frac{B\dot{\alpha}}{U}+K^2H^*_3\alpha+K^2H^*_4\frac{h}{B}	\\
		&-2\tilde{C}_L\chi_{Lu}\frac{u}{U}+(\tilde{C}_L^\prime+\tilde{C}_D)\chi_{Lw}\frac{w}{U}	\Bigg),\\
		M=\frac{1}{2}\rho U^2B^2&\Bigg( KA^*_1\frac{\dot{h}}{U}+KA^*_2\frac{B\dot{\alpha}}{U}+K^2A^*_3\alpha+K^2A^*_4\frac{h}{B}\\
		&-2\tilde{C}_M\chi_{Mu}\frac{u}{U}+\tilde{C}_M^\prime\chi_{Mw}\frac{w}{U}	\Bigg),
	\end{aligned}
\end{equation}
where $H_i^*=H^*(K)$ and $A^*=A^*(K)$ are the lift and moment aerodynamic derivatives, respectively, for $K=2\pi/V_r$ being the motion reduced frequency; $\chi=\chi(K)$ is the aerodynamic admittance for the corresponding wind velocities and forces, for $K$ being gust reduced frequency; and, $\tilde{C}_L^\prime$ and $\tilde{C}_M^\prime$ are the slopes of the static aerodynamic coefficients. \par
In the case of a flat plate, the slopes are $\tilde{C}_L^\prime=2\pi$ and $\tilde{C}_M^\prime=\pi/2$. The aerodynamic derivatives relate to the analytical Theodorsen function (omitted for brevity; see~\cite{simiuWindEffectsStructures1996}). The aerodynamic admittance function is the one of~\cite{searsAspectsNonstationaryAirfoil1941}, which is relevant only for the vertical gust, i.e. $\chi=\chi_{Lw}=\chi_{Mw}$ and $\chi_{Lu}=\chi_{Mu}=0$ (infinitely thin plate). The unsteady aerodynamic forces in~\eqref{eq:FP_Forces} can be written in the time domain as:
\begin{equation}\label{eq:FP_SE_TD}
	\begin{aligned}
		L=&-\frac{1}{2}\rho U^2 B2\pi\Bigg\{\int_{-\infty}^\tau \Phi_{se}(\tau-\tau_1)\left[\alpha_a^\prime(\tau_1)+\frac{h^{\prime\prime}(\tau_1)}{B}+\frac{\alpha^{\prime\prime}(\tau_1)}{4}\right]\text{d}\tau_1\\
		&+\int_{-\infty}^\tau \Phi_{b}(\tau-\tau_1)\frac{w^\prime(\tau_1)}{U}\text{d}\tau_1\Bigg\}-\frac{ \pi}{4} \rho U^2\left({h}^{\prime\prime}+B\alpha^\prime\right),\\
		M=&\frac{1}{2}\rho U^2 B^2\frac{\pi}{2}\Bigg\{\int_{-\infty}^\tau \Phi_{se}(\tau-\tau_1)\left[\alpha_a^\prime(\tau_1)+\frac{h^{\prime\prime}(\tau_1)}{B}+\frac{\alpha^{\prime\prime}(\tau_1)}{4}\right]\text{d}\tau_1\\
	    &+\int_{-\infty}^\tau \Phi_{b}(\tau-\tau_1)\frac{w^\prime(\tau_1)}{U}\text{d}\tau_1\Bigg\}-\frac{\pi}{16} \rho B^2U^2\left(\alpha^\prime+\frac{\alpha^{\prime\prime}}{8}\right),\\
	\end{aligned}
\end{equation}
where $\Phi_{se}=\Phi_{se}(\tau)$ and $\Phi_{b}=\Phi_{b}(\tau)$ are the Wagner and K\"ussner functions, describing the rise-time of the motion-induced forces. We employ~\eqref{eq:FP_SE_TD} for data generation using the approximation of $\Phi_{se}$ by~\cite{jonesUnsteadyLiftFinite1939}, and consider the rise-time of the gust-induced forces in the frequency domain through the approximation of the aerodynamic admittance $\chi$ by~\cite{giesingNonlinearTwodimensionalUnsteady1968}.\par
The objective is to employ the GP model for predicting unsteady aerodynamic coefficients and aeroelastic responses during buffeting and flutter, with the aim of addressing the physics omitted in the QS model. A comparison between the linear unsteady model in Equation~\eqref{eq:FP_SE_TD} and the QS model in Equation~\eqref{eq:QS} reveals that the QS model neglects fluid memory, i.e., $\Phi_{se}=\Phi_{b}=1$, and the apparent mass (last terms in Equation~\eqref{eq:FP_SE_TD}). The aerodynamic center is $n_L=n_M=1/4.$\par

\begin{figure*}[!t]
	\centering
	\includegraphics[clip]{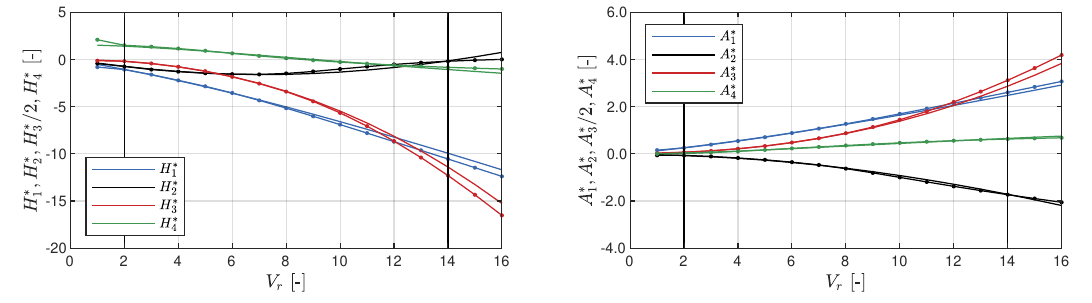} 
	\caption{Flat Plate - Forced Vibration (prediction): Aerodynamic derivatives for analytical (solid line) the GP (solid-dot line) models of the lift (left) and moment (right). The forced sinusoidal input motion is with amplitude of $\alpha_{h0}=\alpha_{a0}=1$ deg. The GP prediction is based on the mean posterior. The black lines indicate the reduced velocity range used to generate the random motion signals used for training ($2\leq V_r\leq 14$ - cf. Fig.~\ref{fig:FP_Train}).}
	\label{fig:FP_FD}
\end{figure*}

\begin{figure*}[!t]
	\centering
	\includegraphics[clip]{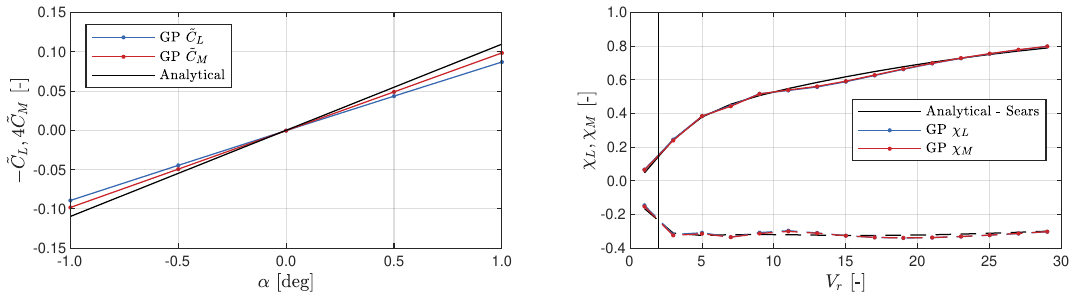} 
	\caption{Flat Plate - Forced Excitation (prediction): Static aerodynamic coefficients (left) obtained for constant input rotation; aerodynamic admittance functions (right) obtained for sinusoidal vertical gust acting on a static section with gust amplitude $\alpha_{w0}=0.5$ deg.}
	\label{fig:FP_SWC_ADM}
\end{figure*}

\begin{table}[t]
	\centering
	\footnotesize
	\begin{tabularx}{0.78\linewidth}{c c c c c}
		\hline
		\multicolumn{5}{c}{Dynamic Properties}  \\[2pt]
		$m_h$ & $m_\alpha$ & $f_h$& $f_\alpha$&$\xi$ \\[2pt]
		[t/m] & [tm$^2$/m]& [Hz] & [Hz] & [$\%$] \\[2pt]			
		\hline
		22.74 & 2470 & 0.100 & 0.278 & 0.3 (FP) / 0.5 (GB) \\[2pt]			
		\hline
	\end{tabularx}
	\caption{Dynamic properties for aeroelastic analyses of a flat plate (FP) and Great Belt's deck section (GB): $m_h$=vertical mass; $m_\alpha$=torsional mass; $f_h$=vertical frequency; $f_\alpha$=torsional frequency; $\xi$=damping ratio.}
	\label{tab:DynParam}
\end{table}

\begin{figure*}[!t]
	\centering
	\includegraphics[clip]{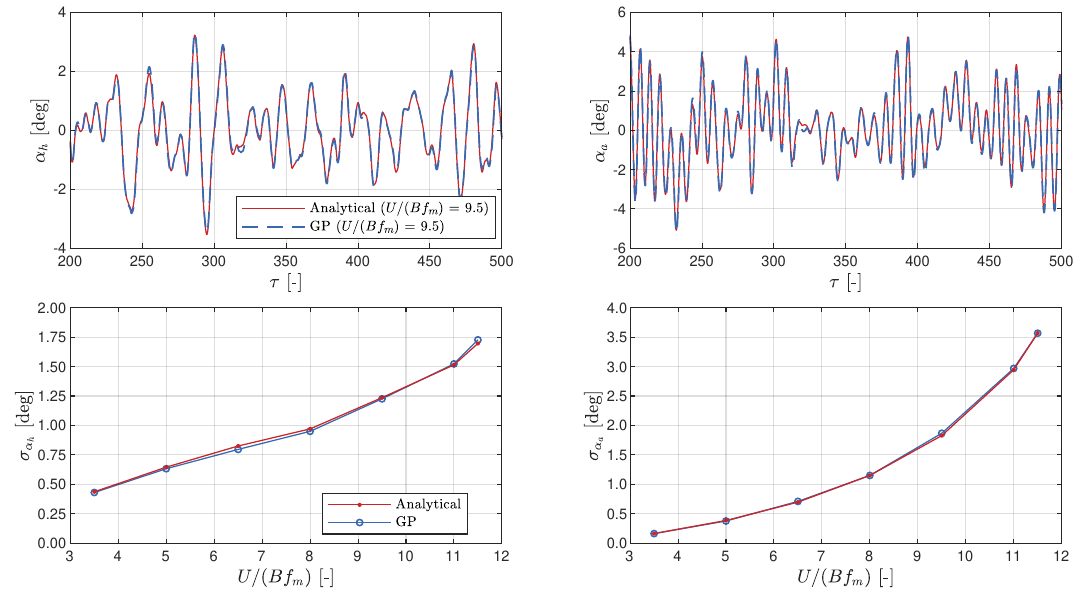} 
	\caption{Flat Plate - Buffeting Analysis (prediction): Sample time histories of the vertical (top-left) and torsional (top-right) displacements for a wind speed of $U/(B/f_M)=9.5$; standard deviation of the vertical (bottom-left) and torsional (bottom-right) displacements for a range of mean wind speeds. The turbulence intensity is $I_w=5$ \%.}
	\label{fig:FP_Buf}
\end{figure*}

\begin{figure*}[!t]
	\centering
	\includegraphics[clip]{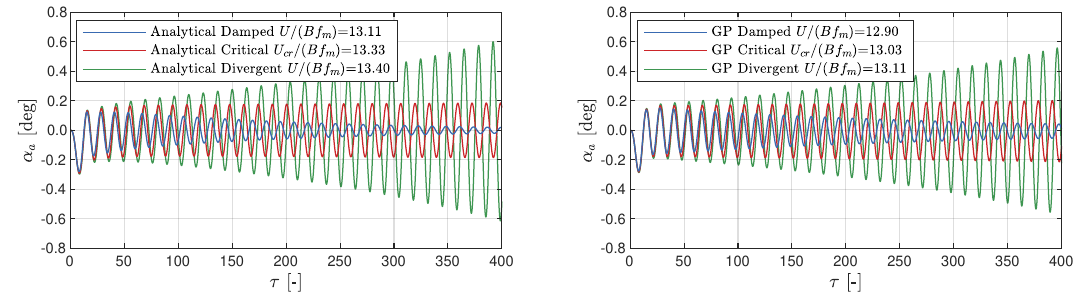} 
	\caption{Flat Plate - Flutter Analysis (prediction): Rotation time histories for the analytical (left) and GP (right) models at a damped, critical flutter ($U_{cr}/(Bf_m)$) and divergent wind speed, with $f_m=(f_h+f_a)/2$ being the central oscillation frequency.}
	\label{fig:FP_Flut_TH}
\end{figure*}

\subsection{Training}

The GP model is trained on the lift and moment due to gusts and motions acting concurrently, based on the training parameters listed in Tab.~\ref{tab:NumParam}. The number of lags is selected such that it covers more than 98 \% of the rise time in the aerodynamic forces, i.e., $\Phi_b \rightarrow 1$ and $\Phi_{se} \rightarrow 1$.\par
Figure~\ref{fig:FP_Train} (top) depicts the training time histories of the motion and gust angles. Two training signals with low and high amplitudes for both motion $\alpha_m$ and gust $\alpha_g$ angles are stacked together to later test buffeting prediction at turbulence intensity different from those in the training set. In other words, the input for each training signal consists of gust and motion angles acting concurrently; yielding the input set of the two signals $\left\{(\alpha_m^l,\alpha_g^l);(\alpha_m^h,\alpha_g^h)\right\}$, where $l$ denotes low and $h$ denotes high motion amplitude/turbulence intensity. A representative training sample of the lift and moment is given in Fig~\ref{fig:FP_Train} (bottom). The GP model fits well the analytical signals with a practically negligible variance due to low noise (the force time-histories were contaminated with a signal-to-noise ratio of 20). The model mismatch between the QS and linear unsteady (analytical) model is captured by the data-driven part of the hybrid GP model.\par 

\subsection{Forced excitation (gust and motion)}

Having the GP model trained, we proceed to verify its prediction of standard aerodynamic coefficients through forced excitation. First, we obtain the aerodynamic derivatives (cf. Fig.~\ref{fig:FP_FD}) through forced sinusoidal vibration for 12 cycles at each reduced velocity, separately for the vertical displacements and rotation with amplitude $\alpha_{h0}=\alpha_{a0}=1$ deg. Excellent correspondence is observed between the analytical and predicted values. Similar results were obtained in~\cite{kavrakovDatadrivenAerodynamicAnalysis2022}; however, an important difference that we verify here is that the presented GP model is capable of predicting pure motion input, despite being trained on mixed gust and motion input.\par 
The GP model also demonstrates robust predictive capabilities even in the $V_r$ range outside the training space. This is evident in the flutter derivatives (cf. Fig.~\ref{fig:FP_FD} outside the black lines) and the static aerodynamic coefficients (cf. Fig.~\ref{fig:FP_SWC_ADM}, left), since the latter represent quasi-steady asymptotic values at $V_r\rightarrow\infty$. The static aerodynamic coefficients are obtained by imposing constant values of $\alpha_a$. The prediction is good and the minor difference is be attributed to the data-driven part requiring to yield zero forces since QS model already incorporates this information.\par
Figure~\ref{fig:FP_SWC_ADM} (right) depicts the complex aerodynamic admittance functions, which are obtained by imposing 12 cycles of sinusoidal gusts with amplitude $\alpha_w=0.5$ deg on a static section. Both the complex and real part of the admittance are in excellent correspondence with the analytical Sears function. The data-driven part of the model is thus capable of separating the effect of the rise time in the buffeting forces, despite being trained on random free-stream turbulence following the von K\'arm\'an spectrum.\par

\begin{figure*}[!t]
	\centering
	\includegraphics{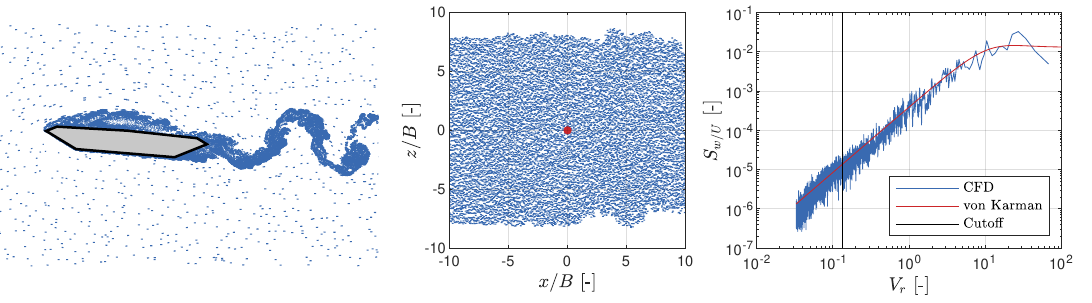} 
	\caption{Great Belt Deck Section: Particle maps of a CFD simulation with a deck section subjected free-stream turbulence (left) and a CFD simulation without a deck section considering only inflow free-stream particles (centre). The spectrum of the vertical gust angle (right) is computed based on the fluctuations at the centre of the CFD domain from a simulation without a section (centre, red dot). The cutoff limit represents the seeding frequency of the inflow particles. }
	\label{fig:GB_PMat}
\end{figure*}
\subsection{ Buffeting and flutter analyses}	

Finally, we perform aeroelastic buffeting and flutter analyses by one-step ahead dynamic time integration and dynamic properties listed in Tab.~\ref{tab:DynParam}. The buffeting analyses are conducted for wind speeds ranging between $3.5\leq U/(Bf_m)\leq11.5$, where $f_m=(f_h+f_a)/2$ is the central oscillation frequency, at each wind speed for a total time of $\tau$=600. The input gust angles are generated using von K\'arm\'an spectrum with turbulence intensity of $I_w=5$ \%. This intensity is not the same as during training but falls within the training range since we trained on two separate signals with $I_w=3$ \% and $I_w=6$ \%. \par 
Figure~\ref{fig:FP_Buf} presents a sample response time history from the buffeting analysis at wind speed of $U/(Bf_m)=9.5$ (top), along with the standard deviation of the response for the selected wind speed range (bottom). Excellent correspondence can be observed for both models, showcasing the robustness of the GP model compared to the typically observed numerical instabilities for data-driven models. The model is capable of predicting turbulence intensities other than the training ones. \par
The flutter analysis is conducted by subjecting the dynamic system to an initial displacement and allowing it to oscillate freely. Figure~\ref{fig:FP_Flut_TH} depicts time histories of the rotation at wind velocities below (damped), at (critical), and above (divergent) the flutter speed. The GP model successfully predicts coupled flutter at the critical speed $U_{cr}=U/(Bf_m)=13.03$, with a negligible difference of 2 \% compared to the analytical value $U_{cr}=U/(Bf_m)=13.33$. For comparison, the QS model (without the apparent mass, as used in the GP prior), resulted in low critical speed $U/(Bf_m)\leq 1$ - a very low wind speed due to positive moment aerodynamic centre and no apparent mass effects. \par

\section{Great Belt Bridge Deck Section}

The Great Belt Bridge is often considered a benchmark in bridge aerodynamics~\citep{larsenAerodynamicAspectsFinal1993,dianaIABSETaskGroup2022}. We employ the presented methodology to showcase the applicability of the GP model in simulating non-linear aerodynamic behavior, predicting aerodynamic coefficients, and conducting aeroelastic analyses of a 2D section of the Great Belt deck. To extract the training data and verify the predictions, we utilize CFD analyses based on the Vortex Particle Method~\citep{cottetVortexMethodsTheory2000} in the GPU implementation of~\cite{morgenthalImmersedInterfaceMethod2007,morgenthalGPUacceleratedPseudo3DVortex2014}, noting that any CFD method can be used to generate training data.\par
The deck geometry corresponds to the H4.1 section of~\cite{reinholdWindTunnelTests1992}, scaled for a chord of $B=31$ m and discretized into 250 panels, resulting in a simulation with a time step of $\Delta\tau_{\mathrm{CFD}}=0.0165$ at a Reynolds number of $\mathrm{Re}=1\times10^5$. The free-stream turbulence is simulated by seeding inflow particles every 4th simulation step, with prescribed turbulence characteristics (see Tab.\ref{tab:NumParam}; \cite{prendergastSimulation2DUnsteady2006,hejlesenEstimatingAerodynamicAdmittance2015}). Similar simulation settings to those in \cite{kavrakovSynergisticStudyCFD2018} are used in this specific example. We refer to this study for a comprehensive CFD validation of the aerodynamic coefficients, verification of free-stream turbulence statistics, and aeroelastic analyses. For illustration, Fig.\ref{fig:GB_PMat} (left) presents a particle map of a buffeting analysis at $U=30$ m/s, while Fig.~\ref{fig:GB_PMat} (center) presents a particle map of the free-stream turbulence from a simulation without an immersed deck. The spectrum of the monitored vertical fluctuations at the center of the CFD domain (red dot) is verified with the target von K\'arm\'an spectrum in Fig.~\ref{fig:GB_PMat} (right).\par

\begin{figure*}[!t]
	\centering
	\includegraphics{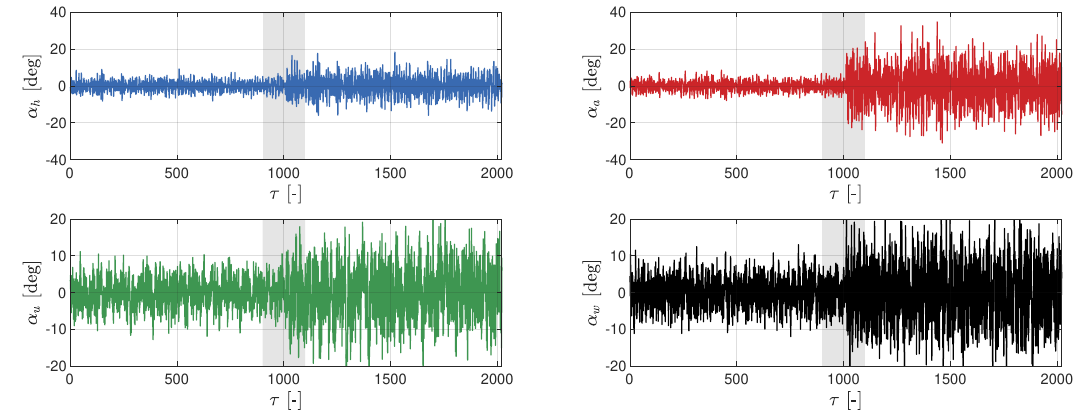} 
	\includegraphics{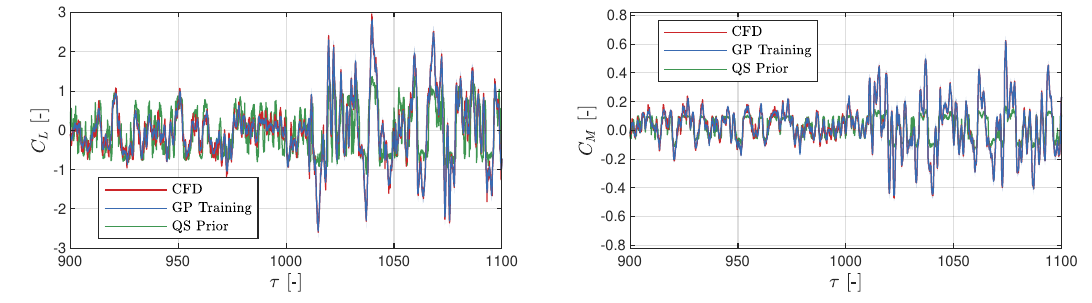} 	
	\caption{Great Belt Deck Section - Forced Excitation (training): Input training signals of the vertical motion (top-left), rotation (top-right), lateral gust (centre-left) and vertical gust (centre-right) angles. Sample training time histories of the aerodynamic lift (bottom-left) and moment (bottom-right) coefficients. These time histories correspond to the shaded region in the input training angles.}
	\label{fig:GB_Train}
\end{figure*}
\begin{figure}[!t]
	\centering
	\includegraphics{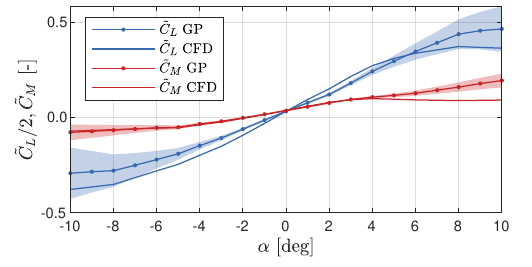} 
	\caption{Great Belt Deck Section - Forced Excitation (prediction): Static aerodynamic coefficients obtained for constant input rotation. The shaded region represents the 99 \% confidence interval of the prediction for the GP model.}
	\label{fig:GB_SWC}
\end{figure}

\subsection{Training}	

In contrast to the flat plate case, the training dataset for the Great Belt Deck is considerably larger as it was shown to be necessary in order for the GP model to capture the intricate aerodynamics. We employ a combination of simulation data that encompasses input motion or gust angles acting concurrently at a mean wind speed of $U=20$ m/s. Specifically, we include motion angles with low $\alpha_a^{l}$ and high $\alpha_m^h$ amplitude (cf. Tab.~\ref{tab:NumParam}), along with gust angles with low $\alpha_g^{l}$ and high $\alpha_g^{h}$ turbulence intensity. The chosen motion angle amplitudes aim to cover both the linear and non-linear amplitude ranges, while the gust angles are set for an isotropic turbulence with 6 \% and 12 \% for the high and low intensity cases, respectively. The gust and motion angles are acting concurrently; yielding the input set of the two stacked signals $\left\{(\alpha_m^l,\alpha_g^l);(\alpha_m^h,\alpha_g^h)\right\}$. This covers a range of cases that enables the GP model to learn, in a single training procedure, the motion-induced and gust-induced forces since they are inseparable in the non-linear range~\citep{tesfayeNumericalInvestigationNonlinear2022}. Augmenting the training set with separate gust and motion input angles, in addition to the case of concurrent gust and motion, could potentially further isolate their individual effects and lead to better predictions. Nevertheless, this aspect was not explored in the present study.\par 
We note that the time-histories for the input gust angles are obtained from a CFD simulation without a deck section, by tracking the lateral $u(t)$ and vertical $w(t)$ fluctuations at the centre of the domain (the red dot in Fig.~\ref{fig:GB_PMat} - centre). Given that the inflow particles are identical for the simulations with and without section, it is assumed that the input gust angles remain the same. This implies minimal effect of the section on the incoming flow and minor viscous effects~\citep{kavrakovSynergisticStudyCFD2018}. We employ similar strategy for predictions involving gust input.\par
\begin{figure*}[!t]
	\centering
	\includegraphics{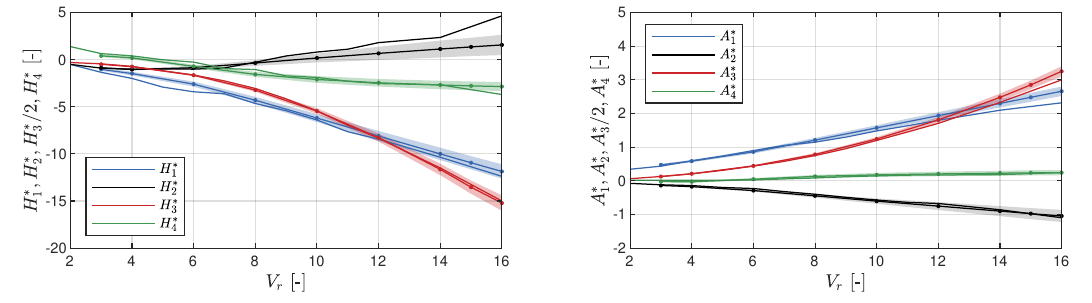} 
	\caption{Great Belt Deck Section - Forced Vibration (prediction): Aerodynamic derivatives for analytical (solid line) the GP (solid-dot line) models of the lift (left) and moment (right). The forced sinusoidal input motion is with amplitude of $\alpha_{h0}=\alpha_{a0}=1$ deg. The shaded area represent the 99 \% confidence interval for the GP model, obtained based on 1000 samples from the harmonic posterior for each reduced velocity.}
	\label{fig:GB_FD}
\end{figure*}

Figure~\ref{fig:GB_Train} depicts a sample of the training signals and the corresponding lift and moment coefficients, showcasing a good fit of the GP model. Additionally, the figure presents the QS prior mean, relying on CFD static aerodynamic coefficients (cf. Fig.~\ref{fig:GB_SWC}) and the aerodynamic centers ($n_L$ and $n_M$) obtained from the CFD flutter derivatives at $V_r=16$ (cf. Fig.~\ref{fig:GB_FD}; see \cite{kavrakovAeroelasticAnalysesBridges2018} for discussion). The QS model  underestimates the amplitude at large angles of attack and lags behind the QS model - typical consequential effects due to neglecting non-linear fluid memory. These effects are captured by the data-driven component of the hybrid GP model.\par

\begin{figure*}[!t]
	\centering
	\includegraphics{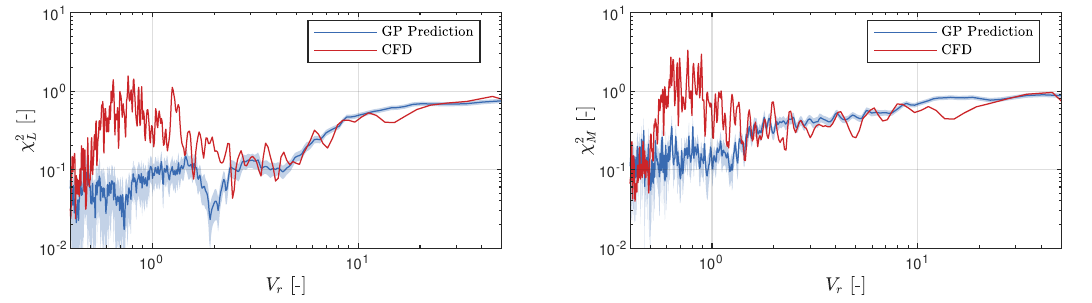} 
	\caption{Great Belt Deck Section - Forced Excitation (prediction): Aerodynamic admittance functions for lift (left) and moment (right). The admittance is obtained from a forced excitation analysis with a random free stream turbulence and a static deck. The input free-stream turbulence for the GP model is based on a CFD simulation without a section deck (cf. Fig.~\ref{fig:GB_PMat}, centre and right), which is also used in the normalization when computing the CFD admittance. The shaded area represent the 99 \% confidence interval for the GP model, obtained based on 1000 samples from the random posterior of the GP model.}
	\label{fig:GB_Adm}
\end{figure*}

\begin{figure*}[!t]
	\centering
	\includegraphics{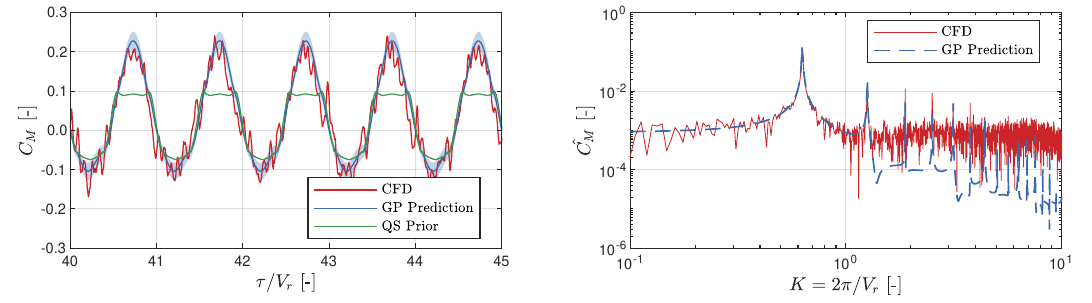} 
	\caption{Great Belt Deck Section - Forced Vibration (prediction): Sample time history of the moment coefficient (left) and its corresponding fast Fourier transform (right) for sinusoidal input rotation with high amplitude $\alpha_{a0}=10$ deg at $V_r=10$.}
	\label{fig:GB_HA}
\end{figure*}

\begin{figure*}[!t]
	\centering
	\includegraphics{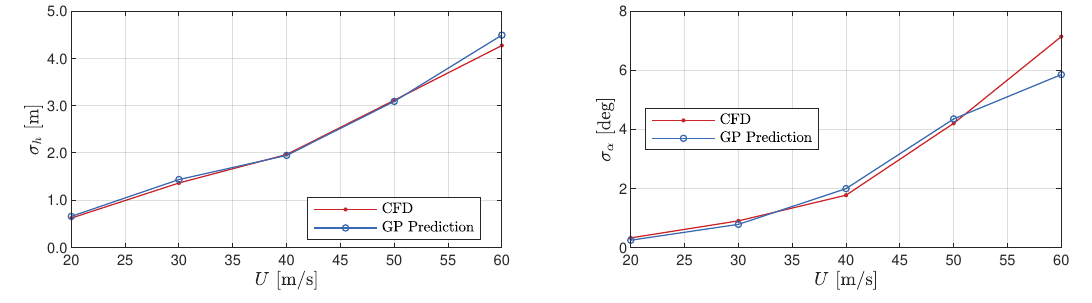} 
	\caption{Great Belt Deck Section - Buffeting Analysis (prediction): Standard deviation of the vertical (left) and torsional (right) displacements for a range of mean wind speed. The turbulence intensity is $I_w=10$ \%.}
	\label{fig:GB_Buf_RMS}
\end{figure*}

\begin{figure*}[!t]
	\centering
	\includegraphics{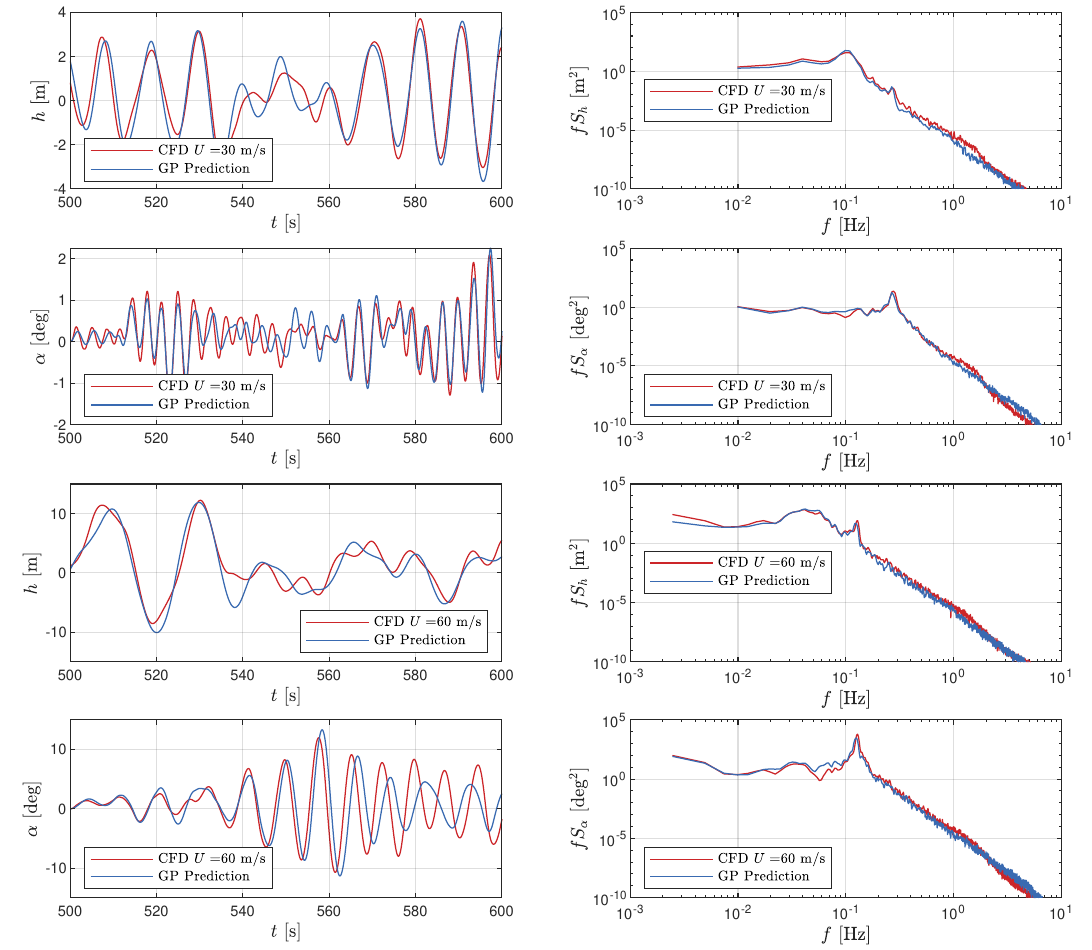} 
	\caption{Great Belt Deck Section - Buffeting Analysis (prediction): Sample time histories of the vertical and torsional displacements at $U=$30 m/s (left-top) and at $U=$60 m/s (left-bottom); corresponding power spectral density of the vertical and torsional displacements (right). The turbulence intensity is $I_w=10$ \%.}
	\label{fig:GB_Buf}
\end{figure*}

\begin{figure*}[!t]
	\centering
	\includegraphics{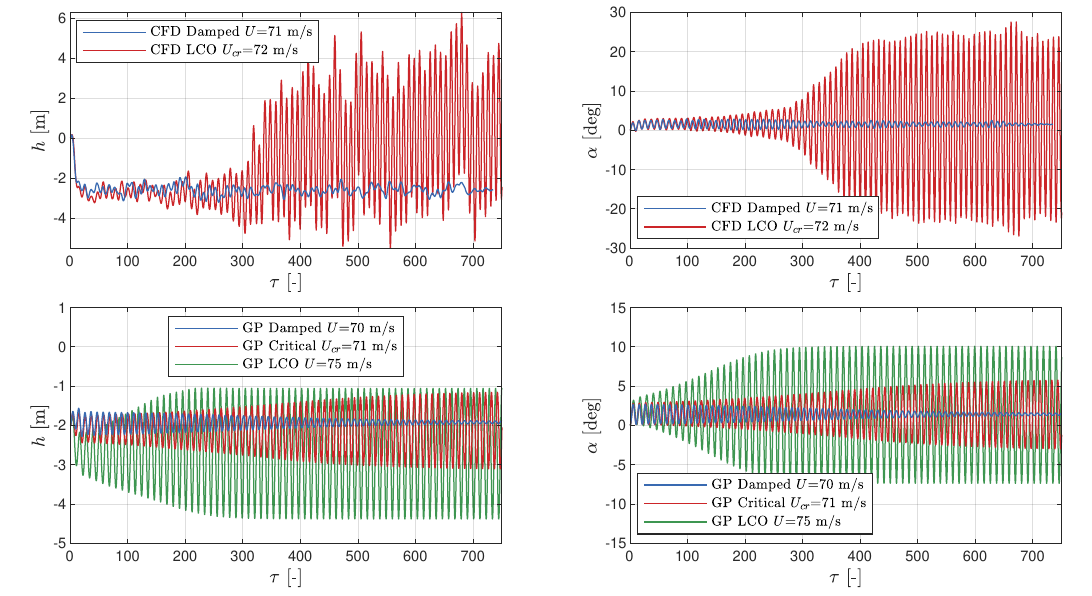} 
	\caption{Great Belt Deck Section - Flutter Analysis (prediction): Time histories of the vertical (left) and torsional (right) displacements for the CFD (top) and GP (bottom) models.}
	\label{fig:GB_F}
\end{figure*}

\subsection{Forced excitation (gust and motion)}

We utilize the model for multi-step ahead prediction for forced excitation. A good way to examine both the model robustness for predictions outside the $V_r$ training range and quasi-steady non-linearity is through the static aerodynamic coefficients. We impose a constant rotation and obtain the static lift and moment coefficient, which are depicted in Figure~\ref{fig:GB_SWC}. Interestingly, there is a difference between the GP and CFD models, despite using the QS prior mean based on the CFD model coefficients. Although the CFD model is near the GP confidence interval, this difference is based on the data-driven part of the GP model. It is conjectured that the information extracted by the GP model from the forced vibration has less noise-to-signal ratio due to vortex shedding than the static CFD simulations. Similar discrepancies were observed beyond $\pm$5 degrees when comparing the CFD coefficients with wind tunnel data from other studies in \cite{kavrakovSynergisticStudyCFD2018}. The GP model remains stable for predictions outside the training range, i.e. $V_r\rightarrow \infty$, and can capture the non-linear trend of the static aerodynamic coefficients.\par
Next, we cover the aerodynamic derivatives to verify the linear range of the motion-induced forces, based on sinusoidal input motion with vertical and rotational amplitudes of $\alpha_{h0}=\alpha_{a0}=1$ degree, for a total of 7 cycles per reduced velocity. Figure~\ref{fig:GB_FD} depicts the aerodynamic derivatives, including their 99 \% confidence range, obtained based on 1000 samples that were sampled from the harmonic force posterior prediction. A good correspondence can be observed, with the CFD model typically being near the confidence interval of the GP model. Similar results were obtained by~\cite{kavrakovDatadrivenAerodynamicAnalysis2022} for a model trained on purely motion-induced input. In contrast to their study, the presented GP model is trained on significantly a larger dataset containing both gust- and motion angles, which are incorporated in the model in a non-linear fashion. Thus, it is important to demonstrate the GP model's capability of isolating motion-induced forces. \par
The linear frequency-dependent gust-induced forces are an integral part of the model. We verify these forces by predicting the aerodynamic admittance functions for a static section and gust input angles based on isotropic free-stream turbulence with intensity of 10 \% and a wind speed of $U=30$ m/s, different from the training set that was based on $U=20$ m/s and turbulence intensities of 6\% and 12\%. Figure~\ref{fig:GB_Adm} depicts the aerodynamic admittance of the CFD and GP model including the 99 \% confidence range, obtained based on 1000 samples that were sampled from the force posterior predictions. A good correspondence can be observed, with the GP model appearing as a moving mean fit to the CFD model. An interesting observation is that the GP model filters out the effect of vortex shedding in the forces, as evidenced by the peak in the CFD admittance in the range of $0.3 \lesssim V_r \lesssim 1$. These forces result from alternating vortices shed in the wake of a static section. The Strouhal number of the current section is between 0.106 and 0.173~\citep{kavrakovSynergisticStudyCFD2018}, which corresponds to the reduced velocity range where the peak appears in the CFD admittance. It is expected that these forces are negligible in buffeting analysis since they are not related to the motion or gust input. As the QS prior mean results in a unit admittance for the whole $V_r$ range, the results demonstrate that the hybrid GP model is capable of capturing the linear fluid memory effects in the gust-induced forces.\par
Non-linear fluid memory in the motion-excited forces is typically manifested through higher-order harmonics for a single-frequency input. We test the model's capability in predicting this type of frequency-dependent non-linearity by forcing the deck to rotate sinusoidally, with a large amplitude of $\alpha_a=10$ deg at reduced velocity of $V_r=10$. Fig.~\ref{fig:GB_HA} (left) depicts the fluctuating moment coefficient, along with its Fourier transform (right). The GP model is capable of isolating higher-order harmonics in the motion-induced forces, despite being trained on a mixed dataset. Interestingly, the QS prior mean predictions significantly underestimate the moment for positive angles of attack, while the GP model renders good results that capture the non-linear frequency-dependent effects.

\begin{figure*}[!t]
	\centering
	\includegraphics{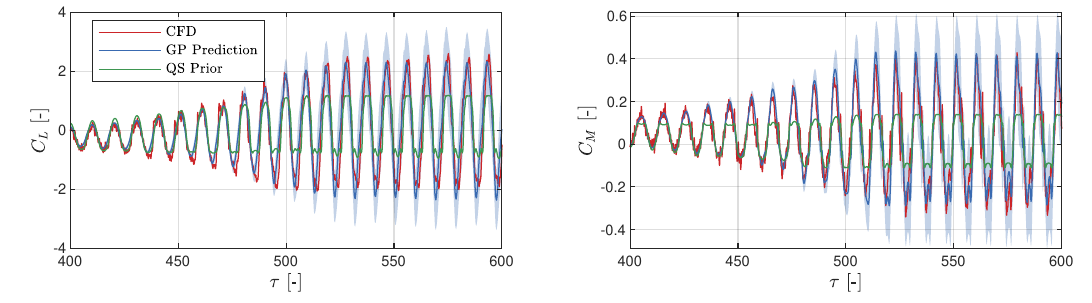} 
	\caption{Great Belt Deck Section - Forced Vibration (prediction): Sample time histories of the lift (left) and moment (right) coefficients based on the CFD displacements during LCO. In case of the GP model, a forced vibration analysis is conducted employing the CFD motion angles as input (see Fig.~\ref{fig:GB_F}, top).}
	\label{fig:GB_FL}
\end{figure*}

\subsection{Buffeting and flutter analyses}

We study the model predictive capabilities for aeroelastic analyses using the dynamic characteristics listed in Tab.~\ref{tab:DynParam}. The buffeting analyses are conducted at six different wind speeds, ranging from $U=20$ m/s up to $U=60$ m/s, each for $t=600$ sec. We used turbulence intensity of 10 \%, which is the range between the training turbulence intensities of 6 \%  and 12 \%. The input gust angles for the GP model are tracked at the centre of the domain of a section-less simulation to facilitate comparative basis between the CFD and GP models, as discussed previously. \par
Figure~\ref{fig:GB_Buf_RMS} depicts standard deviation of displacements for the selected wind speed range, while Fig.~\ref{fig:GB_Buf} depicts a sample time histories of the vertical and rotational displacements (left) and the Welch spectra of the complete time histories (right) at $U=30$ m/s and $U=60$ m/s. Generally, the GP model yields good predictions in the standard deviation for both vertical and torsional responses across wind speeds, with some discrepancy at $U=60$ for the torsional response. This can be also observed in the time histories; however, it should be noted that the oscillations are very large. Thus, the tracking point of the wind fluctuations that serve as an input is different to what the CFD section experiences. Apart from this reason, it is difficult to pindown other causes for the discrepancy. It could be conjectured that there is non-stationary amplitude modulation for the CFD model, which is difficult to be captured by the GP model without introducing autoregressive terms.\par 
Finally, we conduct flutter analysis and predict the critical flutter velocity and LCOs under laminar free-stream. Figure~\ref{fig:GB_F} depicts time histories from the flutter analysis. The critical wind speed is $U_{cr}=71$ m/s for the GP and $U_{cr}=72$ m/s for the CFD model. This small discrepancy was not observed if the training is conducted based on motion input only \cite{kavrakovDatadrivenAerodynamicAnalysis2022}; thus, the higher noise due to the turbulent forces can be a plausible reason.\par
Further, the GP model underestimates the amplitudes of the LCO, showing a gradual increase in the response amplitudes — a soft flutter,  in contrast to the CFD results that indicating a hard flutter through violent change in amplitude. The static component in the vertical displacements is slightly underestimated by the GP model. However, this underestimation is minor due to the overall good agreement in the static coefficients (cf. Fig.~\ref{fig:GB_SWC}). To examine the discrepancies, we perform forced vibration using the LCO displacements from the CFD analysis (see Fig.~\ref{fig:GB_F}, top) as input for the GP model, and compare the resulting GP forces with the CFD forces during the LCO (see Fig.~\ref{fig:GB_FL}). It can be observed that the CFD forces falls within the uncertainty interval of the GP forces; the QS prior mean underestimates the amplitudes significantly - a discrepancy addressed by the data-driven part of the hybrid GP model. Similar results were obtained for the GP model of the motion-induced forces in~\cite{kavrakovDatadrivenAerodynamicAnalysis2022}, where it was argued that capturing hard flutter is challenging without modeling a self-regressed form of the flutter dynamics due to non-stationary frequency modulation~\citep{amandoleseLowSpeedFlutter2013}. Soft flutter amplitudes were successfully captured for an H-shaped deck in \cite{kavrakovDatadrivenAerodynamicAnalysis2022}; it is conjectured that the present hybrid GP model possess the same capability.

\section{Conclusions}

We presented a hybrid methodology for modeling the gust- and motion-induced forces acting on bluff bodies using data-driven GPs and a semi-analytical QS model as a prior mean. The hybrid non-linear model, constructed in an entirely non-dimensional aerodynamic form, takes gust- and motion-induced angles as input, aiming to address model errors resulting from assumptions in the semi-analytical prior. A bespoke procedure was devised for training, using forced motion and free-stream turbulence concurrently. The methodology was utilized to predict the aerodynamic forces due to forced excitation and aeroelastic response during flutter and buffeting. Verification included testing on flat plate linear aerodynamics and the non-linear aerodynamics of a 2D section of the Great Belt Bridge deck.\par

Despite being trained on concurrent motion- and gust-induced training data, the hybrid GP model effectively predicted Theodrsen's aerodynamic derivatives and Sears' complex admittance function. This showcased its ability to distinguish between motion- and gust-induced components within the data. The model's robustness in handling broadband input in an aeroelastic setting was demonstrated through accurate predictions of buffeting and flutter responses for linear flat plate aerodynamics.\par

We demonstrated the predictive capabilities of the hybrid model in bridge aerodynamics based on CFD training data. This data contains high interior noise caused by vortex shedding and body-induced turbulence, as well as their interaction with free-stream turbulence. The model successfully predicted both linear aerodynamics, confirmed through verification of the frequency-dependent coefficients, and non-linear aerodynamics, characterized by higher-order harmonics.\par 

The good correspondence between CFD and GP buffeting responses demonstrated the new applicability of the presented methodology. We observed a small discrepancy in the standard deviation of the rotation at high wind speeds. The critical flutter velocity showed good agreement between the CFD and GP models; however, the GP model could not capture the LCO amplitudes during hard flutter. Nevertheless, the non-linear flutter forces were accurately captured when using CFD LCOs as input for the GP model. It is conjectured that the discrepancies in the buffeting analysis at high wind speeds and the inability to capture LCO amplitudes is attributed to the complex non-stationary flutter dynamics. Further development is necessary to address such intricate hard flutter dynamics, although the practical usability of predicting hard flutter LCOs in civil structures remains questionable.\par

Several aspects of the hybrid methodology warrant improvements, including incorporating better semi-analytical aerodynamic priors, selecting more appropriate covariance functions, employing sparse GPs to reduce training time, and designing input signals for gust-induced angles based on wind tunnel experiments. For example, considering the hybrid non-linear unsteady model instead of the QS model as an aerodynamic prior could yield better predictions~\citep{kavrakovSynergisticStudyCFD2018}, since less model uncertainty would need to be captured by the data-driven part. The effect of training data quality, such as numerical/experimental uncertainty or interior noise from  vortex shedding, needs to be further addressed as well.\par 

A straightforward extension of the hybrid model is to include drag forces and lateral motion as inputs, facilitating 3D multimode buffeting and flutter analyses based on the strip assumption.  Although the span-wise coherence of the wind fluctuations can be considered, as they represent an exogenous input to the GP model at each strip, the limitation of the strip assumption regarding 3D admittance persists. Utilizing the presented methodology to study the non-linear interaction between gust- and motion-induced forces, investigate the influence of free-stream turbulence on flutter (in a 2D sense), or explore online learning based on monitoring data remain viable prospects.

\appendix

\section{Covariance Function}\label{App:CovFunc}

The exponential covariance function with Automatic Relevance Detection (ARD) reads:
\begin{equation}\label{eq:Kernel}
		k(\boldsymbol{\alpha}_i,\boldsymbol{\alpha}_j)=a^2\exp\left(-\frac{1}{2}(\boldsymbol{\alpha}_i-\boldsymbol{\alpha}_j)\boldsymbol{Q}(\boldsymbol{\alpha}_i-\boldsymbol{\alpha}_j){^T}\right),
\end{equation}
with variance $a^2$ with $\boldsymbol{Q}=\mathrm{diag}(\boldsymbol{l}^{-2})$, where $\boldsymbol{l}$ is the length scale vector containing the individual length scales
\begin{equation}\label{eq:Kernel2}
		\boldsymbol{l}=(l_{h^\prime},l_{a^\prime},l_{u^\prime},l_{w^\prime},l_{h,i},l_{a,i},l_{u,i},l_{w,i},\dots l_{h,i-S},l_{a,i-S},l_{u,i-S},l_{w,i-S}).
\end{equation}
The ARD property facilitates the independent scaling of each dimension by multiplying the matrix $\boldsymbol{Q}$, of size $\boldsymbol{Q}\in\mathbb{R}^{8+4S\times 8+4S}$, with the input $\boldsymbol{\alpha}\in\mathbb{N}^{N_s\times8+4S}$. For example, the input term related to the vertical velocity for the training steps $N_s$, $\boldsymbol{\alpha}_{h}^\prime\in\mathbb{N}^{N_s\times 1}$, is scaled through its separate length scale $l_{h^\prime}$.

\section*{Data Statement}

The Matlab code, including the results from Sec.~\ref{sec:FundApp}, is publicly available online: \href{github.com/IgorKavrakov/AeroGPBuff}{\color{forceblue}github.com/IgorKavrakov/AeroGPBuff}.

\section*{Author contributions}

Igor Kavrakov: Conceptualization, Methodology, Software, Funding acquisition, Writing - Original Draft. Guido Morgenthal: Conceptualization, Software, Writing - Review. 	Allan McRobie: Conceptualization, Methodology, Supervision, Writing - Review. 

\section*{Acknowledgments}

IK gratefully acknowledges the support by the German
Research Foundation (DFG) [Project No. 491258960], Darwin College and the Department of Engineering, University of Cambridge. GM gratefully acknowledges the support by the German
Research Foundation (DFG) [Project No. MO 2283/14-1].

\bibliographystyle{elsarticle-harv}
\bibliography{references}

\end{document}